\documentclass[aps,prd,preprint,showpacs,groupaddress]{revtex4}

\begin{document}

\title{Sterile Neutrinos as the Warm Dark Matter in the Type II Seesaw Model}

\author{\footnotesize Wan-lei Guo\footnote{Email address: guowl@itp.ac.cn} }

\address{Kavli Institute for Theoretical Physics China,
Institute of Theoretical Physics, \\ Chinese Academy of Science,
Beijing 100080, P.R.China}

\begin{abstract}
In the framework of type II seesaw mechanism we discuss the number
of sterile right-handed Majorana neutrinos being the warm dark
matter (WDM). When the type II seesaw mass term $M_\nu ^{II}$ is far
less than the type I seesaw mass term $M_\nu ^{I}$, only one of
three sterile neutrinos may be the WDM particle. On the contrary,
the WDM particles may contain all sterile neutrinos. If  $M_\nu
^{II} \sim M_\nu ^{I}$, the allowed number is not more than ${\cal
N} - 1$ for ${\cal N}$ sterile neutrinos. It is worthwhile to stress
that three different types of neutrino mass spectrum are permitted
when $M_\nu ^{II} \gg M_\nu ^{I}$ and $M_\nu ^{II} \sim M_\nu ^{I}$.
\end{abstract}


\maketitle

\newpage

\section{Introduction}

Recent solar \cite{SNO}, atmospheric\cite{SK}, reactor\cite{KM} and
accelerator\cite{K2K} neutrino oscillation experiments have provided
us with very robust evidence that neutrinos are massive and lepton
flavors are mixed. The ordinary type I seesaw mechanism
\cite{SEESAW} gives a very simple and appealing explanation of the
smallness of left-handed neutrino masses -- it is attributed to the
largeness of right-handed neutrino masses. On the other hand, recent
cosmological observations have provided convincing evidence in favor
of the existence of Dark Matter (DM) \cite{DM}. To clarify the
identity of the DM remains a prime open problem in particle physics
and cosmology. The idea that right-handed Majorana neutrinos may be
the Warm Dark Matter (WDM) has been investigated in detail
\cite{WDM}. It has been shown that a sterile neutrino with the mass
of a few keV appears to be a viable warm dark matter candidate in
the $\nu$ Minimal Standard Model ($\nu$MSM) \cite{VMSM}. This model
is very interesting since it can also explain neutrino oscillations,
baryon asymmetry \cite{BA}, inflation\cite{Inflation}, the observed
velocities of pulsars \cite{Pulsar} and the early reionization
\cite{Reionization}. In the $\nu$MSM, only one of three right-handed
sterile neutrinos can be the WDM particle.

If there is an additional ${\rm SU(2)_{\rm L}}$ Higgs triplet, one
can derive the so-called type II seesaw mechanism\cite{T2}. The
neutrino mass matrix $M_\nu$ is composed of two parts. Hence we may
relax the WDM constraints on the parameters of the $\nu$MSM. In this
note, we shall investigate the WDM in the type II seesaw model. The
remaining part of this paper is organized as follows. In Section II,
we briefly describe the main features of the type II seesaw model.
In Section III, constrains from various cosmological observations
are shown for the WDM. In Section IV, we shall discuss the number of
sterile neutrinos being the WDM in detail. Finally the summary are
given in Section V.

\section{The type II seesaw model}

In the type II seesaw model, the Lagrangian relevant for neutrino
masses reads \cite{T2,GUO}:
\begin{eqnarray}
-{\cal L} & = & \frac{1}{2} \overline{N^c_{\rm R}}  M_{\rm R} N_{\rm
R} + M^2_\Delta {\rm Tr}(\Delta_{\rm L}^\dagger \Delta_{\rm L})
+ \overline{\psi_{\rm L}}\; Y_\nu N_{\rm R} H  \nonumber \\
& & + \overline{\psi_{\rm L}^c}\; Y_\Delta i \tau_2 \Delta_{\rm L}
\psi_{\rm L} - \mu H^T i \tau_2 \Delta_{\rm L} H + h.c.  \; ,
%       (1)
\end{eqnarray}
where $\psi_{\rm L}= (\nu_{\rm L}$, $l_{\rm L})^T$  and
$H=(H^0,H^-)^T$ denote the left-handed lepton doublet and the
Higgs-boson weak isodoublet respectively, $N_{\rm R}$ stands for the
sterile right-handed Majorana neutrino singlets, and
\begin{equation}
\Delta_{\rm L}= \left( \matrix{ \frac{1}{\sqrt{2}}\Delta^+ &
\Delta^{++} \cr \Delta^0 & - \frac{1}{\sqrt{2}} \Delta^+ }\right)
%       (2)
\end{equation}
is the ${\rm SU(2)_L}$ Higgs triplet. After spontaneous gauge
symmetry breaking, we have
\begin{eqnarray}
-{\cal L} & = & \frac{1}{2} \left ( \overline{\nu_{\rm L}},
\overline{N_{\rm R}^c} \right ) \left ( \matrix{ M_{\rm L} & M_{\rm
D} \cr M_{\rm D}^{ T} & M_{\rm R} \cr } \right ) \left (\matrix{
\nu_{\rm L}^c \cr N_{\rm R} \cr } \right ) + h.c.   \; .
%       (3)
\end{eqnarray}
Then one may obtain the effective (light and left-handed) neutrino
mass matrix $M_\nu$ via the type II seesaw mechanism:
\begin{equation}
M_\nu \; = M_{\rm  L} - M_{\rm  D} M^{-1}_{\rm  R} M^T_{\rm  D} \; =
M_\nu ^{II}+ M_\nu ^{I} \;
%       (4)
\end{equation}
where $M_{\rm D} \equiv Y_\nu \langle H \rangle$ with $\langle H
\rangle = v \simeq 174 {\rm GeV}$ and $M_{\rm  L} \equiv 2 Y_\Delta
\langle \Delta^0 \rangle$ with $\langle \Delta^0 \rangle
 \simeq \mu^\ast v^2/ M^2_\Delta$.
$M_\nu ^{I}= - M_{\rm  D} M^{-1}_{\rm  R} M^T_{\rm  D}$ is the
ordinary type I seesaw mass term, and the type II seesaw mass term
$M_\nu ^{II}=M_{\rm L}$ arises from the additional Higgs triplet
vacuum expectation value. Without loss of generality, both $M_{\rm
R}$ and the charged lepton mass matrix $M_l$ can be taken to be
diagonal, real and positive; i.e., $ M_{\rm R} = {\rm
Diag}\{M_1,M_2,M_3\}$ with $M_1 \leq M_2 \leq M_3$ and $ M_{l} =
{\rm Diag}\{m_e,m_\mu,m_\tau\}$.  The flavor eigenstates $\nu_{\rm
L}$ can be expressed as $\nu_{\rm L} = K \hat{\nu}_{\rm L}+ R
\hat{N}^c_{\rm R}$, where $R \approx M_{\rm D} M_{\rm R}^{-1}$ and
$K$ is an approximate unitary matrix. $N = \hat{N}_{\rm R} +
\hat{N}^c_{\rm R}$ and $\nu = \hat{\nu}_{\rm L} + \hat{\nu}^c_{\rm
L}$ are the heavy and light Majorana neutrino mass eigenstates,
respectively.

The type I seesaw mass term $M_\nu ^{I} $ can be diagonalized as
follow:
\begin{eqnarray}
{M_\nu ^{I}}^{diag}  = {\rm Diag} \{
\bar{m}_1,{\bar{m}_2},{\bar{m}_3 }\} = - U^\dagger M_{\rm  D}
M^{-1}_{\rm  R} M^T_{\rm  D} U^* = - [S_1 + S_2 + S_3] \;
%       (5)
\end{eqnarray}
where  $S_I$ denotes a contribution from each sterile neutrino $N_I$
and is given by $(S_I){_{ij}} = X_{iI}X_{jI}$ with $X_{iI} =
(U^\dagger M_{\rm D})_{iI}/\sqrt{M_I}$. When $M_\nu ^{II} =0$,
${M_\nu ^{I}}^{diag}$ and $U$ are the diagonal neutrino mass matrix
and the Maki-Nakagawa-Sakata (MNS) lepton flavor mixing
matrix\cite{MNS}, respectively.  Since ${\rm Det} [{M_\nu
^{I}}^{diag} + S_i ] = 0$, one may arrive at
\begin{eqnarray}
\bar{m}_1  \bar{m}_2 X_{3i}^2 + \bar{m}_1  \bar{m}_3 X_{2i}^2
+\bar{m}_2  \bar{m}_3 X_{1i}^2 +\bar{m}_1 \bar{m}_2  \bar{m}_3 =0 \;
.
%       (6)
\end{eqnarray}
By taking the trace of both sides in Eq.(5), we find that
\begin{eqnarray}
\bar{m}_1 + \bar{m}_2 + \bar{m}_3 = \left| - \sum_{i=1}^3 (X_{i1}^2
+ X_{i2}^2 +  X_{i3}^2)\right| \leq \sum_{i=1}^3 (|X_{i1}|^2 +
|X_{i2}|^2 + |X_{i3}|^2) \; .
%       (7)
\end{eqnarray}

\section{Constraints on the warm dark matter}

In this scenario, there are not any stable particles. When the
active-sterile mixing matrix $R$ is sufficiently small, the lifetime
of sterile neutrinos ($M_I \ll m_e$) will exceed the age of the
universe. These sterile neutrinos may be the WDM particles. The
production mechanism of  sterile neutrinos is due to the
active-sterile neutrino oscillations \cite{WDM}. In terms of the
correct dark matter density, one can derive\cite{VMSM}
\begin{eqnarray}
\sum_I \sum_{i=1,2,3} \left| {M_{\rm  D}}_{iI} \right| = m_0^2  \; ,
%       (8)
\end{eqnarray}
where $m_0 = {\cal{O}}( 0.1 ) {\rm eV}$ and the summation of $I$ is
taken over the sterile neutrino $N_I$ being dark matter. The above
equation can be reexpressed as
\begin{eqnarray}
\sum_I \sum_{i=1,2,3} \frac{M_I}{M_1} \left| {X}_{iI} \right|
=\frac{ m_0^2}{M_1} \equiv m_\nu^{dm}\; .
%       (9)
\end{eqnarray}
The sterile neutrino masses $M_I$ can receive constraints from
various cosmological observations and the possible mass range is
very restricted as \cite{0612182}
\begin{eqnarray}
0.3 \; {\rm keV} < M_I < 3.5 \;  {\rm keV} \; ,
%       (10)
\end{eqnarray}
where the lower bound is given by the Tremaine-Gunn bound\cite{TGB},
while the upper bound is given by the radiative decays of sterile
neutrinos in dark matter halos limited by X-ray
observations\cite{0612182}. The stronger constraint coming from the
Lyman-$\alpha$ observations \cite{Lyman} is $M_I \geq 10 \; {\rm
keV}$ which is inconsistent with  Eq.(10). Therefore, if the
Lyman-$\alpha$ constraint is taken for granted, the production of
sterile neutrinos due to active-sterile neutrino transitions happens
to be too small to account for observed abundance of dark matter. In
other words, physics beyond our model is likely to be required to
produce dark matter sterile neutrinos \cite{Inflation,Pulsar}.
Another option is to assume that the universe contained relatively
large lepton asymmetries\cite{LA}.

\section{Warm dark matter in the type II seesaw model}

In the type I seesaw model, only one of three right-handed neutrinos
can be the WDM particle \cite{VMSM}. Since the neutrino mass matrix
$M_\nu$ contains two parts of contributions in our scenario,
interesting results can be obtained. In the following parts, we
shall discuss the number of sterile neutrinos being the WDM in terms
of Eqs.(9) and (10). When $M_\nu ^{II} \ll M_\nu ^{I}$, we can
derive the same conclusions as in Ref.\cite{VMSM}. If $M_\nu ^{II}
\gg M_\nu ^{I}$, it is obvious that all right-handed neutrinos may
be the WDM particles.

In this section, we shall analyze the $M_\nu ^{II} \sim M_\nu ^{I}$
case in detail. It is worthwhile to stress that all three kinds of
neutrino mass spectrum (Normal hierarchy, Inverted hierarchy and
Degenerate) are permitted when $M_\nu ^{II} \gg M_\nu ^{I}$ and
$M_\nu ^{II} \sim M_\nu ^{I}$. Moreover, one can derive the same
consequences for different neutrino mass spectrum. For illustration,
we assume that the neutrino mass spectrum is the normal hierarchy
case, i.e., $m_1 < m_2 \ll m_3$ with $m_3 \approx \sqrt{\Delta
m_{\rm atm}^2} = 0.05 \, {\rm eV}$ \cite{FIT}. When $M_\nu ^{II}
\sim M_\nu ^{I}$, one can directly obtain $\bar{m}_3 \sim m_3 $
where we have assumed $   \bar{m}_1 \leq \bar{m}_2 \leq \bar{m}_3$.
$\bar{m}_1$ and $ \bar{m}_2$ may be  equal to zero.

If there are only two heavy Majorana neutrinos $N_1$ and $N_2$,
$\bar{m}_1 = 0$ holds \cite{IJMPE}. When both sterile neutrinos are
assumed to be the dark matter,  Eq.(7) becomes
\begin{eqnarray}
\bar{m}_2 + \bar{m}_3 \leq \sum_{i=1}^3 (|X_{i1}|^2 + |X_{i2}|^2)
\leq m_\nu^{dm} \; ,
%       (11)
\end{eqnarray}
where we have used  Eq.(9). The above inequality can not be
satisfied since $\bar{m}_3 \sim 0.05 \, {\rm eV}$ and $m_\nu^{dm}
\sim {\cal} 10^{-5} \, {\rm eV}$.

When only one  of two sterile neutrinos, say  $N_1$, is assumed to
be the dark matter, one can directly derive $|X_{11}|^2 + |X_{21}|^2
+ |X_{31}|^2 = m_\nu^{dm}$ from  Eq.(9). Since $\bar{m}_1 = 0$,
Eq.(6) induces $X_{11}^2 \bar{m}_2 \bar{m}_3 =X_{12}^2 \bar{m}_2
\bar{m}_3 =0$ which implies $\bar{m}_2 =0$ and (or)
$X_{11}=X_{12}=0$. For the $\bar{m}_2 =0$ case, one may deduce
\begin{eqnarray}
{M_{\rm  D}}_{i1} \propto {M_{\rm  D}}_{i2} \; (i=1,2,3)
%       (12)
\end{eqnarray}
from  Eq.(5). When $X_{11}=X_{12}=0$, the first row and column of
$S_1$ and $S_2$ vanish. Then Eq.(5) is reduced to that for $2 \times
2$ matrices:
\begin{eqnarray}
{\rm Diag} \{{\bar{m}_2},{\bar{m}_3 }\} + X_{i1} X_{j1} = - X_{i2}
X_{j2} \; (i,j =2,3) \; .
%       (13)
\end{eqnarray}
The vanishing determinant leads to ${\bar{m}_2}{\bar{m}_3 } +
{\bar{m}_2} X_{31}^2 + {\bar{m}_3} X_{21}^2 =0$. The upper bound of
${\bar{m}_2}$ turn out to be
\begin{eqnarray}
{\bar{m}_2} = \left| X_{21}^2 +
\frac{{\bar{m}_2}}{{\bar{m}_3}}X_{31}^2 \right| \leq  |X_{21}|^2 +
|X_{31}|^2 \leq m_\nu^{dm} \; .
%       (14)
\end{eqnarray}
Therefore,  one of two sterile neutrinos may be the dark matter in
the type II seesaw model.

Now, let us discuss the case including three heavy neutrinos. It is
obvious that all three sterile neutrinos can not simultaneously be
the WDM from Eqs.(7) and (9). If two of three sterile neutrinos, say
$N_1$ and $N_2$, are the WDM particles. Making use of the (2,3)
block of Eq.(5), we have
\begin{eqnarray}
\left|({\bar{m}_2} + X_{21}^2 + X_{22}^2){\bar{m}_3} \right|\approx
\left|(X_{21}X_{31}+X_{22}X_{32}) \right|^2 \leq \frac{1}{4}
{m_\nu^{dm}}^2 \; .
%       (16)
\end{eqnarray}
Hence we can derive ${\bar{m}_2} \lesssim m_\nu^{dm}$ since
$|X_{21}|^2 + |X_{22}|^2 \leq m_\nu^{dm}$ and $\bar{m}_3 \sim 0.05
\, {\rm eV}$. When ${\bar{m}_1}={\bar{m}_2}=0$, one may also arrive
at
\begin{eqnarray}
{M_{\rm  D}}_{i1} \propto {M_{\rm  D}}_{i2} \propto {M_{\rm D}}_{i3}
\; (i=1,2,3) \; .
%       (15)
\end{eqnarray}
For the ${\bar{m}_2}\neq0$ case, we can obtain
\begin{eqnarray}
\bar{m}_1 = \left|  X_{11}^2 + \frac{\bar{m}_1}{\bar{m}_2} X_{21}^2
+ \frac{\bar{m}_1}{\bar{m}_3} X_{31}^2\right| \leq |X_{11}|^2 +
|X_{21}|^2 +|X_{31}|^2 \leq {m_\nu^{dm}}
%       (17)
\end{eqnarray}
with the help of Eqs.(6) and (9).

Finally, we consider the remaining possibility that only one sterile
neutrino, say $N_1$, plays a dark matter particle. For the
${\bar{m}_1}= {\bar{m}_2}= 0$ case, the  Eq.(16) can also be
obtained. When ${\bar{m}_1}= 0$, Eq.(6) induces $X_{11} = X_{12} =
X_{13} = 0$. If  ${\bar{m}_1}\neq{\bar{m}_2}\neq0$, we can derive
the same conclusion as in Ref.\cite{VMSM}: $\bar{m}_1 \leq
m_\nu^{dm}$.

\section{Summary}

We have analyzed the number of sterile neutrinos that can explain
the warm dark matter in the type II seesaw model. When $M_\nu ^{II}
\ll M_\nu ^{I}$, only one of three right-handed sterile neutrinos
may be the WDM particle \cite{VMSM}. If $M_\nu ^{II} \gg M_\nu
^{I}$, the WDM particles may contain all sterile neutrinos. In this
note, the $M_\nu ^{II} \sim M_\nu ^{I}$ case is detailed discussed.
We find that the allowed number is not more than ${\cal N} - 1$ for
${\cal N}$ sterile neutrinos. It is worthwhile to stress that three
different types of neutrino mass spectrum are permitted when $M_\nu
^{II} \gg M_\nu ^{I}$ and $M_\nu ^{II} \sim M_\nu ^{I}$.

\acknowledgments{This work was supported by the China Postdoctoral
Science Foundation, the K. C. Wong Education Foundation (Hong Kong)
and in part by the National Nature Science Foundation of China
(NSFC) under the grant 10475105 and 10491306.}

\end{document}